\documentclass[prd,aps,a4paper,floatfix,amsmath,amssymb,twocolumn]{revtex4-2}
\usepackage[margin=3cm]{geometry}
\usepackage{mathtools}
\usepackage{graphicx}
\usepackage{array}
\usepackage[dvipsnames]{xcolor}
\usepackage[toc,page]{appendix}
\usepackage{soul}
\usepackage[pdftex,breaklinks,colorlinks,
linkcolor=Blue,
citecolor=teal,
anchorcolor=red,
urlcolor=cyan]{hyperref}

\newcommand{\mus}{\mu}
\newcommand{\mys}{s}
\newcommand{\Omegas}{\tilde \Omega}

\begin{document}
%%%%%%%%%%%%%%%%%%%%%%%%%%%%%%%%%%%%%%%%%%%%%%%%%%%%%%%%%%%%%

\title{Critical collapse of an axisymmetric ultrarelativistic 
	fluid in $2+1$ dimensions}

\author{Patrick Bourg, Carsten Gundlach}
\affiliation{Mathematical Sciences, University of Southampton,
	Southampton SO17 1BJ, United Kingdom}
\date{04 October 2021, revised 15 March 2022}

\begin{abstract}
	We carry out numerical simulations of the gravitational
	collapse of a rotating perfect fluid with the
	ultrarelativistic equation of state $P=\kappa\rho$, in
	axisymmetry in $2+1$ spacetime dimensions with $\Lambda<0$.
	We show that for $\kappa \lesssim 0.42$, the critical
	phenomena are type~I, and the critical solution is stationary.
	The picture for $\kappa \gtrsim 0.43$ is more delicate: for
	small angular momenta, we find type~II phenomena, and the
	critical solution is quasistationary, contracting
	adiabatically. The spin-to-mass ratio of the critical
	solution \textit{increases} as it contracts, and hence, so does
	that of the black hole created at the end as we fine-tune
	to the black-hole threshold. Forming extremal black holes is
	avoided because the contraction of the critical
	solution smoothly ends as extremality is approached.
\end{abstract}

\maketitle

\tableofcontents

%%%%%%%%%%%%%%%%%%%%%%%%%%%%%%%%%%%%%%%%%%%%%%%%%%%%%%%%%%%%%
\section{Introduction}
%%%%%%%%%%%%%%%%%%%%%%%%%%%%%%%%%%%%%%%%%%%%%%%%%%%%%%%%%%%%%

Critical collapse is concerned with the threshold of black-hole
formation in the space of initial data. Starting with Choptuik's study
of the spherically symmetric, massless scalar field \cite{Choptuik93},
and since then generalized to many other systems \cite{GundlachLRR07},
one can enumerate several general features of critical phenomena. The
more interesting kind is now called ``type~II'' phenomena. As the
black-hole threshold is approached through the fine-tuning of a
one-parameter family of initial data, the black-hole mass and
spacetime curvature scale as a power of distance to the black-hole
threshold. Furthermore, for initial data close to the black-hole
threshold, the solution will be, in an intermediary stage, well
approximated by a \textit{critical solution}. This critical solution
has the general characteristics of being self-similar, universal
(independent of the initial data) and possessing a single growing
linear perturbation mode. By contrast, in ``type~I'' phenomena, the
mass and curvature do not scale, but instead approach a nontrivial
constant. The critical solution is time periodic or stationary. The
above properties for type~I and II phenomena hold for all matter
systems studied thus far in $3+1$ and all higher dimensions; see
Ref.~\cite{GundlachLRR07} for a review.

The vast majority of research studies dedicated to critical collapse
focuses on spherically symmetric initial data. Since black holes can
carry angular momentum, the full picture of critical collapse requires
us to go beyond spherical symmetry. However, the generalization
from spherically symmetric to, say, axisymmetric initial data
brings about many numerical and theoretical complications. In part,
one has to deal with an additional independent variable. Moreover,
gravitational waves exist beyond spherical symmetry, and they can
exhibit (type~II) critical phenomena by themselves \cite{Abraham93}.
In particular, this implies that, if one wishes to study the critical
phenomena of some matter field in axisymmetry, one will have to
disentangle the critical phenomena due to the matter field and those
due to gravitational waves. This additional difficulty is currently of
great importance since the critical phenomena of pure gravitational
waves are still poorly understood.

One way to work around those problems is to consider, as a toy model,
the situation in $2+1$ dimensions. Gravity in $2+1$ dimensions is rather
peculiar. Notably, black holes
cannot form without a negative cosmological constant. The rotating
black-hole solution, called the BTZ solution \cite{Banados92}, is also
quite unique. Its central singularity is not a usual curvature
singularity, but a causal singularity. Furthermore, the black-hole
spectrum is separated from the background (anti-de Sitter spacetime)
configuration by a mass gap. A direct consequence of this is that
small deviations from anti-de Sitter spacetime (from now, AdS) cannot
collapse into a black hole. Furthermore, gravitational waves do not
exist in $2+1$ dimensions. For these reasons, it is unclear if many of the results in $2+1$ dimensions can simply be carried over to the $3+1$ dimensional setting. On the other hand, the absence of gravitational waves enables us to bypass the current major difficulties encountered in higher dimensions. A rotating object in $3+1$ dimensions cannot be spherically symmetric, but at most axisymmetric, roughly speaking because rotation tries to flatten the rotating body. In the $2+1$-dimensional counterpart of such rotating axisymmetric solutions, there is no dimension for the flattening to happen, in and all the variables are still only functions of ``time'' and ``radius''. This makes the study of critical collapse with rotating initial data much more tractable.
As in Refs~\cite{Jalmuzna17, Bourg21}, we call a
solution circularly symmetric if it admits a spacelike Killing vector
$\partial_\theta$ with closed orbits. More specifically (and perhaps in a slight abuse of notation), we call it spherically symmetric if there is no rotation, and we call it axisymmetric with rotation.

Pretorius and Choptuik investigated the black-hole threshold for the
spherically symmetric, massless scalar field in $2+1$ dimensions
\cite{Pretorius00}. They found type~II phenomena (mass and curvature
scaling). The critical solution is also found, near the center, to be
approximately \textit{continuously} self-similar (as opposed to
discretely as is the case in $3+1$ dimensions). This system was
investigated in more depth in Ref.~\cite{Jalmuzna15}. The authors
found good agreement between the numerical solution during the
critical regime and an exact continuously self-similar $\Lambda=0$
solution. However, one major unresolved issue is that this exact
solution has \textit{three} growing modes. Investigating the modes
numerically, they surprisingly only find numerical evidence of the top
(largest) growing mode. There is to date no satisfactory explanation
for this, but it was conjectured that the nonlinear effect of the
cosmological constant may be responsible for removing all but the top
growing mode (adding the effect of the cosmological constant
perturbatively does not alter the perturbation spectrum).

In Ref.~\cite{Jalmuzna17}, the same authors generalized the
consideration to a complex rotating scalar field. It turns out that
the effect of rotation is highly nontrivial; neither the critical
exponents nor the critical solution are universal. The angular
momentum does not show any scaling. The mass and curvature may or may
not scale, depending on the one-parameter family of initial
data. Finally, the threshold of mass and curvature scalings are
different.

In Ref.~\cite{Bourg21}, the present authors investigated the
spherically symmetric perfect fluid, with barotropic equation of
state $P=\kappa \rho$. We found that the critical phenomena are of
type~I if $\kappa \lesssim 0.42$, while they are of type~II if $\kappa
\gtrsim 0.43$. The critical solution for type~I is static (as
expected), but for type~II, it is \textit{not} self-similar, but
instead \textit{quasistatic}. That is, the critical solution
corresponds to an adiabatic sequence of static solutions whose size
shrinks to zero (exponentially).

In this paper, we now extend this previous work to
axisymmetric initial data. In Sec.~\ref{section:section2}, we give a
quick overview of the Einstein-fluid matter system. In
Sec.~\ref{section:section3}, we present and discuss our numerical
findings.

%%%%%%%%%%%%%%%%%%%%%%%%%%%%%%%%%%%%%%%%%%%%%%%%%%%%%%%%%%%%%
\section{Einstein and Fluid equations in polar-radial coordinates}
\label{section:section2}
%%%%%%%%%%%%%%%%%%%%%%%%%%%%%%%%%%%%%%%%%%%%%%%%%%%%%%%%%%%%%

We refer the reader to the companion paper \cite{Gundlach21} for a complete
discussion. We use units where $c=G=1$.

In axisymmetry in $2+1$ dimensions, we introduce generalized
polar-radial coordinates as
\begin{eqnarray}
\label{trmetric}
ds^2&=&-\alpha^2(t,r)\,dt^2+a^2(t,r)R'^2(r)\,dr^2 \nonumber \\ &&
+ R^2(r) [ d\theta + \beta(t,r) dt ]^2. 
\end{eqnarray}
We denote $\partial/\partial t$ by a dot and $\partial/\partial r$ by
a prime.
Note that our choice $g_{rr}=a^2R'^2$ makes $a$ invariant under a
redefinition of the radial coordinate, $r\to \tilde r(r)$. The
``area'' (circumference) radius $R$ is defined geometrically as the
length of the Killing vector $\partial/\partial\theta$.

We impose the gauge condition $\alpha(t,0)=1$ ($t$ is proper
time at the center), and the regularity condition $a(t,0)=1$
(no conical singularity at the center).

We define the auxiliary quantity
\begin{equation}
\label{gammadef}
\gamma:=\beta',
\end{equation}
anticipating that $\beta$ will not appear undifferentiated in the Einstein
or fluid equations, but only in the form of $\gamma$ and its derivatives,
since the form (\ref{trmetric}) of the metric is invariant under the change
of angular variable $\theta \to \theta + f(t)$.
It follows that the particular choice of gauge for $\beta$
does not affect our evolution, and for our numerical implementation, we choose $\beta(t,0)=0$. The gauge is fully specified only after specifying the function
$R(r)$. In our numerical simulations, we use the compactified coordinate
\begin{equation}
\label{compactifiedR}
R(r)=\ell\tan(r/\ell),
\end{equation}
where
\begin{equation}
\ell:={1 \over \sqrt{-\Lambda}}
\end{equation}
is the AdS length scale, but for clarity we write $R$ and $R'$ rather than the explicit expressions.

In our coordinates, the Kodama mass $M$ and angular momentum $J$ are given by
\begin{eqnarray}
\label{MJdefcoords}
J(t,r) &:=& {R^3 \gamma \over R' a \alpha},\\
M(t,r)&:=&{R^2\over\ell^2}-{1\over a^2} + {J^2 \over 4 R^2}.
\end{eqnarray}
This local mass function generalizes the well-known Misner-Sharp mass
from spherical symmetry (in any spacetime dimension) to axisymmetry
(in $2+1$ only) \cite{Gundlach21, Kinoshita21}.

The stress-energy tensor for a perfect fluid is 
\begin{equation}
T_{ab}=(\rho+P)u_{a}u_{b}+Pg_{ab},
\end{equation}
where $u^a$ is tangential to the fluid worldlines, with $u^a u_a=-1$, and $P$
and $\rho$ are the pressure and total energy density measured in the fluid frame.
In the following, we assume the one-parameter family of ultrarelativistic fluid
equations of state $P=\kappa\rho$, where $0<\kappa<1$. 

The 3-velocity is decomposed as
\begin{equation}
u^\mu=\{u^t,u^r,u^\theta\}=\Gamma\left\{\frac{1}{\alpha},\frac{v}{aR'},
{w \over R}-\frac{\beta}{\alpha } \right\},
\end{equation}
where $v$ and $w$ are the physical radial and tangential velocities of the fluid
relative to observers at constant $R$, satisfying $v^2 + w^2<1$, and 
\begin{equation}
\Gamma:=\left(1-v^2-w^2\right)^{-1/2}
\end{equation}
is the corresponding Lorentz factor.

The stress-energy conservation law $\nabla_a T^{ab}=0$, which
together with the equation of state governs the fluid evolution,
can be written in balance law form
\begin{equation}
{\bf q}_{,t}+{\bf f}_{,r}={\bf S}, \label{2p1:XY_cons_law}
\end{equation}
where we have defined the conserved quantities
\begin{equation}
{\bf q}:=\{\Omega,Y,Z\}
\end{equation}
given by
\begin{eqnarray}
Y&:=& R'v \sigma,\\
Z &:=& a R^2 R' w \sigma,\\
\Omega&:=& R'R \tau + {J Z \over 2 R^2},
\end{eqnarray}
the corresponding fluxes ${\bf f}$ given by
\begin{eqnarray}
f_{(Y)}&:=& {\alpha\over a}(P+v^2\sigma),\\
f_{(Z)} &:=& \alpha R^2 v w \sigma,\\
f_{(\Omega)}&:=& {\alpha\over a} R v \sigma + {J f_{(Z)} \over 2 R^2},
\end{eqnarray}
the corresponding sources ${\bf S}$ given by 
\begin{eqnarray}
\label{SY}
S_{(Y)}&=& {1\over a}\Bigl[
(w^2-v^2)\sigma\alpha {R'\over R} -\tau\alpha_{,r} \nonumber \\ &&
-(P+v^2\sigma)\alpha(\ln a)_{,r} \nonumber \\ &&
+Rw\sigma\gamma -2v\sigma R' a_{,t} \Bigr], \\
\label{SZ}
S_{(Z)}&=&0, \\
S_{(\Omega)}&=&0,
\label{SOmega}
\end{eqnarray}
and the shorthands
\begin{eqnarray}
\label{sigmadef}
\sigma&:=&\Gamma ^2 (1+\kappa) \rho, \\
P&:=&\kappa\rho, \\
\tau&:=&\sigma-P.
\label{taudef}
\end{eqnarray}

At any given time, the balance laws \eqref{2p1:XY_cons_law} are used
to compute time derivatives of the conserved quantities ${\bf q}$,
using standard high-resolution shock-capturing methods. The ${\bf q}$
are evolved to the next time step via a fourth-order Runge-Kutta
step. At each (sub-)time step, the metric variables are then updated
through the Einstein equations
\begin{eqnarray}
\label{dMdr}
M_{,r} &=& 16 \pi \Omega,\\
\label{dJdr}
J_{,r} &=& 16 \pi Z,\\
\label{dlnalphadr}
(\ln \alpha a)_{,r} &=& 8\pi a^2RR'(1+v^2)\sigma.
\end{eqnarray}

Our numerical scheme is totally constrained, in the sense that only
the matter is updated through evolution equations. Our numerical
scheme exploits the conservation laws for $\Omega$ and $Z$, and
in consequence for $M$ and $J$, to make their numerical
counterparts exactly conserved in the discretized equations.

%%%%%%%%%%%%%%%%%%%%%%%%%%%%%%%%%%%%%%
\section{Numerical results}
\label{section:section3}
%%%%%%%%%%%%%%%%%%%%%%%%%%%%%%%%%%%%%%

%%%%%%%%%%%%%%%%%%%%%%%%%%%%%%%%%%%%%%
\subsection{Initial data}
%%%%%%%%%%%%%%%%%%%%%%%%%%%%%%%%%%%%%%

The numerical grid is equally spaced in the compactified coordinate
$r$, as defined in \eqref{compactifiedR}, with $800$ grid points. We
fix the cosmological constant to be $\Lambda = -\pi^2/4$, so that the
AdS boundary is located at $r = 1$. The numerical domain does not
comprise the entire spacetime. Instead, we set an unphysical outer
boundary with ``copy boundary conditions'' for the conserved variables
${\bf q}$. Unless otherwise stated, we fix the numerical outer
boundary at $r=0.7$, corresponding in area radius to $R_\text{max}
\simeq 1.25 \simeq 1.96 \ell$.

We choose to initialize the primitive fluid variables $\rho$ and $v$ as
double Gaussians in $R$,
\begin{eqnarray}
\rho(0,R) &=& \frac{p_\rho}{2} \left(e^{-\left(
			\frac{R-R_\rho}{\sigma_\rho}\right)^2} +
	e^{-\left(\frac{R+R_\rho}{\sigma_\rho}\right)^2}\right), \nonumber \\ \\
v(0,R) &=& \frac{p_v}{2} \left(e^{-\left(
			\frac{R-R_v}{\sigma_v}\right)^2} + e^{-\left(
			\frac{R+R_v}{\sigma_v}\right)^2}\right),
\end{eqnarray}
where $p_\rho$ and $p_v$ are the magnitudes, $R_\rho$ and
$R_v$ the displacements from the center, and $\sigma_\rho$ and
$\sigma_v$ the widths of the Gaussians. The initial data for $w$ are defined through
the combination $w \Gamma^2$, by
\begin{equation}
w(0,R)\Gamma^2 = p_w R.
\end{equation}
Near the center, the fluid is ``rigidly rotating'' in the sense that
$w \sim R$. The strength of the rotation is parametrized by $p_w$.

As for the spherically symmetric case, we consider three types
of initial data:

1) time-symmetric off-centered: $p_v = 0$, $R_\rho = 0.4$, 

2) time-symmetric centered: $p_v = 0$, $R_\rho = 0$, and 

3) initially ingoing off-centered: $p_v = -0.15$, $R_v = 0.4$, $\sigma_v=0.15$,
$R_\rho = 0.4$.

In all cases, for $\kappa \leq 0.42$, we take $\sigma_\rho = 0.05$,
while for $\kappa \geq 0.43$, we choose $\sigma_\rho = 0.2$. The
reason for this choice is the fact that in spherical symmetry, where
type~I behavior occurs, numerical instabilities form near the
numerical outer boundary and travel inwards. These instabilities form
strong shocks near the center for a second-order limiter. Making the
initial data more compact partially mitigates this. Furthermore, for
$\kappa \leq 0.42$, we use a first-order (Godunov) limiter, as we did
in spherical symmetry. The dissipative properties of the Godunov
limiter eliminate those shocks. As we see, similar instabilities
now also occur for $\kappa \gtrsim 0.43$, for ``large'' deviations
from spherical symmetry. Unless otherwise stated, we therefore also
use a Godunov limiter for $\kappa \geq 0.43$, although we keep
the width as $\sigma_\rho=0.2$.
 
Finally, in all cases, $p_w$ is fixed to a particular value, and
$p_\rho =: p$ is the parameter to be fine-tuned to the
black-hole threshold.

We use the same conventions as in \cite{Bourg21} and denote by
$p=p_\star$ the critical parameter separating initial data that
(promptly) collapse and disperse. We are interested
in initial data where $p \simeq p_\star$, and we denote by
sub$n$ subcritical data for which $\log_{10} (p_\star-p) \simeq
-n$, and by super$n$ supercritical data with $\log_{10}
(p-p_\star) \simeq -n$.

In the following, ``apparent horizon'' (AH) mass and angular momentum
refer to the first appearance of a marginally outer-trapped surface in
our time slicing, indicated by diverging metric component $a$.

%%%%%%%%%%%%%%%%%%%%%%%%%%%%%%%%%%%%%%
\subsection{$\kappa\lesssim 0.42$: Type~I critical collapse}
%%%%%%%%%%%%%%%%%%%%%%%%%%%%%%%%%%%%%%

%%%%%%%%%%%%%%%%%%%%%%%%%%%%%%%%%%%%%%
\subsubsection{Lifetime scaling}
%%%%%%%%%%%%%%%%%%%%%%%%%%%%%%%%%%%%%%

In the spherically symmetric case, we showed in Ref.~\cite{Bourg21}
that the critical phenomena depend on the value of $\kappa$. In
particular, for $\kappa \lesssim 0.42$, we find typical type~I
behavior, where the mass and curvature do not scale and the critical
solution near the center is static. How is this picture modified in
the presence of angular momentum?

\begin{figure*}
	\includegraphics[width=2.\columnwidth, height=1.5\columnwidth]
	{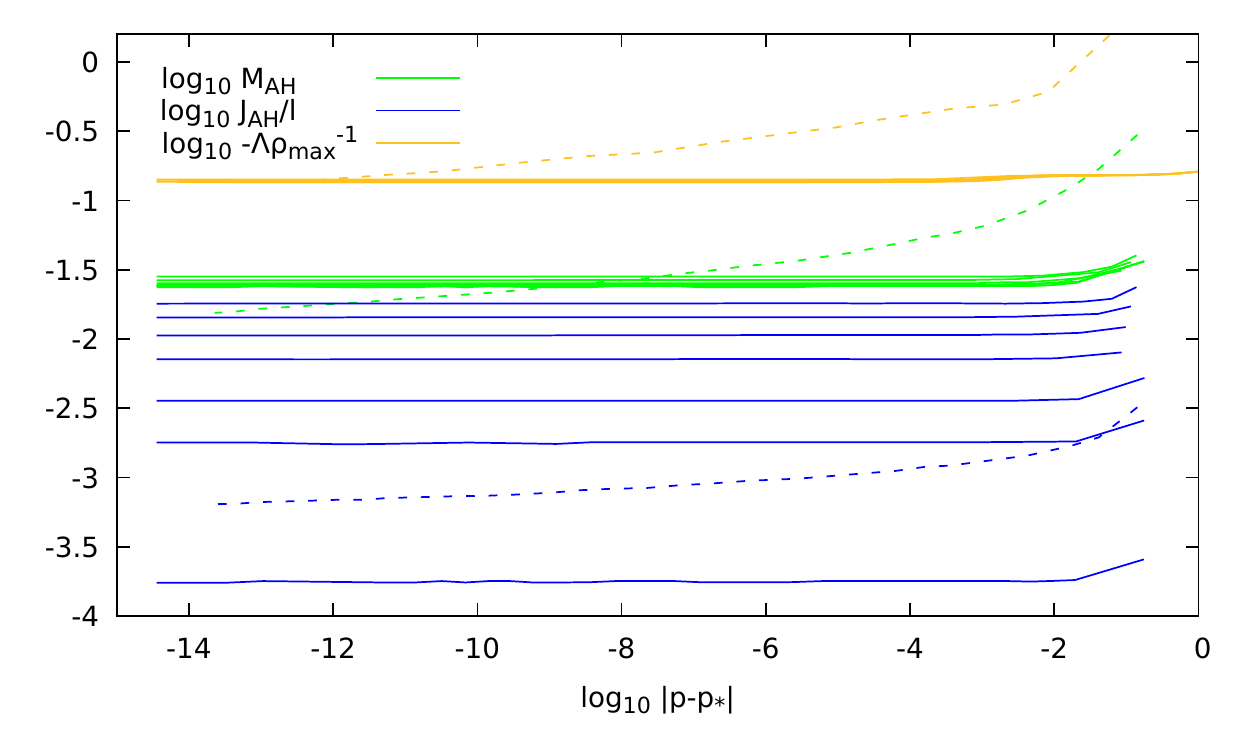}
	\caption{Log-log plot demonstrating the absence of scaling,
		for small $\kappa$, of $\rho_\text{max}$ (upper
		group of curves, orange), $M_\text{AH}$ (middle group of
		curves, green ) and $J_\text{AH}/\ell$ (bottom group of
		curves, blue ) for different values of $p_\omega$, corresponding to
		a range of ``small'' to ``large'' angular momenta. Solid
		lines correspond to $\kappa=0.4$. We find typical type~I
		behavior, irrespective of $p_\omega$. This behavior still
		holds up to $\kappa=0.42$. For comparison, the dotted curves
		show $\kappa = 0.43$ and ``small'' angular momentum, where
		type~II behavior is instead observed within this range of
		fine-tuning, see Fig.~\ref{fig:typeII_powerlaw} for more
		details.}
	\label{fig:typeI_powerlaw}
\end{figure*}

We start answering this question by showing, in
Fig.~\ref{fig:typeI_powerlaw}, a log-log plot of the apparent horizon
mass $M_\text{AH}$ (green group of curves), the
adimensionalized apparent horizon angular momentum
$J_\text{AH}/\ell$ (blue group of curves), and the maximum of
curvature $-\Lambda \rho_\text{max}^{-1}$ (orange group of curves),
against $p-p_\star$, for different values of $p_\omega$. For all these
plots, we consider centered initial data with $\kappa=0.4$. The range
of $p_\omega$ is $p_\omega = 0.01$, $0.1$, $0.2$, $0.4$, $0.6$, $0.8$
and $1.0$. This corresponds to a range from ``small'' angular
momentum (in the sense that $J_\text{AH}/(M_\text{AH}\ell) \ll 1$) to
``large'' angular momentum (for example, $p_\omega = 1.0$ corresponds
to $J_\text{AH}/(M_\text{AH}\ell) \simeq 0.64$).

This behavior, as in spherical symmetry, persists up to some critical
value of $\kappa$ between $\kappa=0.42$ and $\kappa=0.43$. For
comparison, we have also added an evolution with $\kappa=0.43$ (dotted
curves) and $p_\omega=0.01$. The type~I behavior also holds for the
off-centered and ingoing initial data.

% Fig 1 %%%%%%%%%%%%%%%%%%%%%%%%%%%%%%%%%%%%%%%%%%%%%%%%%%%%%%%%%%%%%%%%%
%%%%%%%%%%%%%%%%%%%%%%%%%%%%%%%%%%%%%%%%%%%%%%%%%%%%%%%%%%%%%%%%%%%%%%%%

As in the spherically symmetric case, we consider the
time-dependent quantity $R_M$, defined by
\begin{equation}
M(t,R_M(t)) := 0,
\end{equation}
as a measure of the length scale of the solution.
Similarly, we define the central density $\rho_0$, mass $M_\text{OB}$, and
angular momentum $J_\text{OB}$ at the numerical outer boundary,
\begin{eqnarray}
\rho_0(t) &:=& \rho(t,0),\\
M_\text{OB}(t) &:=& M(t,R_\text{max}),\\
J_\text{OB}(t) &:=& J(t,R_\text{max}).
\end{eqnarray}

\begin{figure*}
	\includegraphics[width=2.\columnwidth, height=1.5\columnwidth]
	{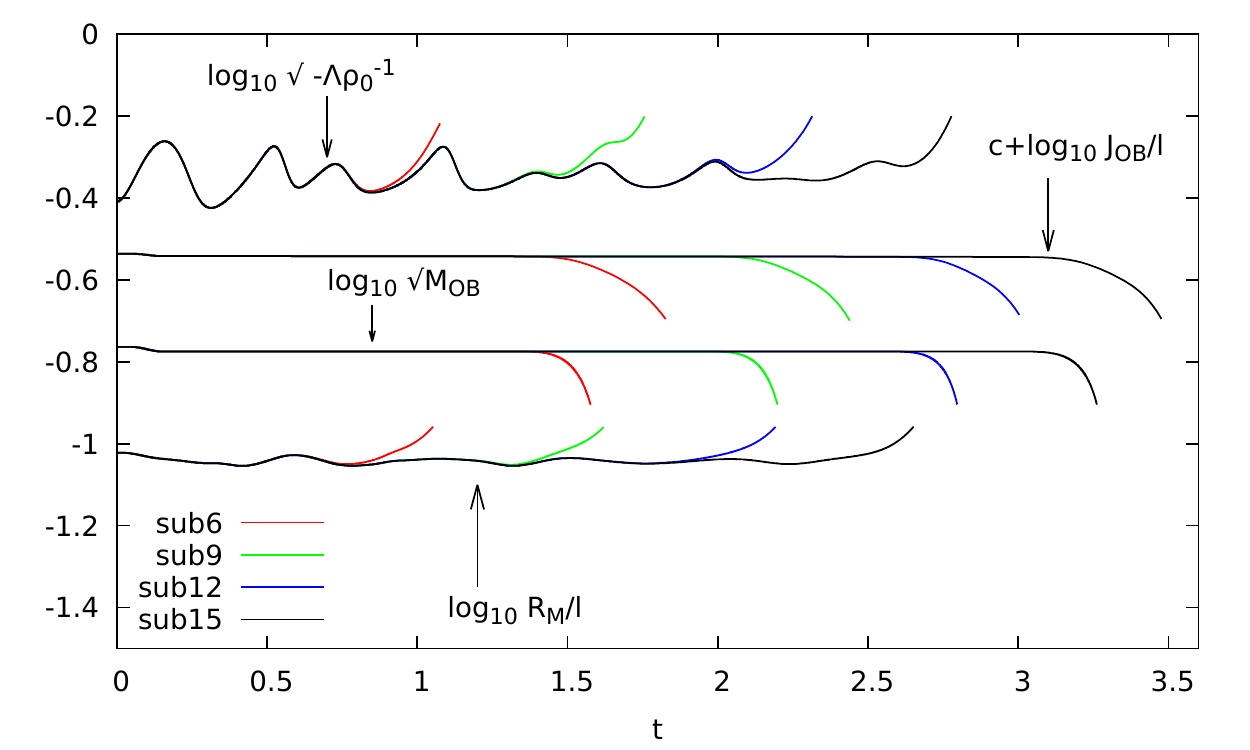}
	\caption{Lin-log plot of $R_M(t)/\ell$, $\sqrt{-\Lambda
		\rho_0^{-1}(t)}$, $\sqrt{M_\text{OB}(t)}$, and
		$J_\text{OB}(t)/\ell+c$ with $c=1.2$, for sub6 to sub15
		centered initial data with $p_w=1.0$. We observe that, as
		we fine-tune to the black-hole threshold, the solution
		approaches an intermediate attractor solution in which all
		the variables are approximately constant. Less fine-tuned
		initial data peel off from the critical regime sooner than
		more fine-tuned data.}
	\label{fig:rho0_JOB_MOB_RM_typeI_sub}
\end{figure*}

In Fig.~\ref{fig:rho0_JOB_MOB_RM_typeI_sub}, we plot in a linear-log
plot, $\sqrt{-\Lambda \rho_0^{-1}}$, $J_\text{OB}/\ell$,
$\sqrt{M_\text{OB}}$, and $R_M/\ell$ against central proper time $t$,
for sub6 to sub15 initial data. We consider here centered initial data
with $p_\omega=1.0$. Note that we have shifted $\log_{10}
J_\text{OB}/\ell$ by a constant $c=1.2$ for clarity. As is
typical for critical phenomena, the more fine-tuned the initial data
to the black-hole threshold, the longer the critical regime, before
the growing mode of the critical solution becomes dominant. In order
to avoid cluttering, we truncate the plots after it is clear that the
growing mode causes the curves to ``peel off'' from the critical
regime.

% Fig 2 %%%%%%%%%%%%%%%%%%%%%%%%%%%%%%%%%%%%%%%%%%%%%%%%%%%%%%%%%%%%%%%%%
%%%%%%%%%%%%%%%%%%%%%%%%%%%%%%%%%%%%%%%%%%%%%%%%%%%%%%%%%%%%%%%%%%%%%%%%

For type~I phenomena, it is not the mass and curvature that scale. Instead, it
is the lifetime of the intermediate regime where the solution is approximated by the critical solution. 
Assuming that the critical solution is stationary, we can make the
following ansatz for its linear perturbations:
\begin{equation}
\delta Z(t,R) = \sum_{i=0}^{\infty} C_i(p)\, e^{\sigma_i {t \over \ell}}\, Z_i(R),
\end{equation}
where $Z$ stands for any dimensionless metric or matter variable, such
as $R^2\rho$, $M$, or $\alpha$.

We assume the critical solution has a single growing mode,
${\rm Re}\, \sigma_0 > 0$. Since the solution is exactly critical at
$p=p_\star$, this implies that $C_0(p) \sim p-p_\star$.
We define the time $t = t_p$ to be the time where the growing perturbation
becomes nonlinear. This occurs when
\begin{equation}
(p-p_\star) e^{\sigma_0 {t_p \over \ell}} \simeq 1,
\label{lifetime}
\end{equation}
and so
\begin{equation}
\label{2p1_ss_crit:lifetimescaling}
t_p = \frac{\ell}{\sigma_0} \ln|p-p_\star| + \text{constant}.
\end{equation}
The exponent $\sigma_0$ can be read off from
Fig.~\ref{fig:rho0_JOB_MOB_RM_typeI_sub}. We find $\sigma_0 \simeq
8.29$, close to its value in spherical symmetry (which was $\simeq
8.84$) \cite{Bourg21}.

%%%%%%%%%%%%%%%%%%%%%%%%%%%%%%%%%%%%%%
\subsubsection{The critical solution}
%%%%%%%%%%%%%%%%%%%%%%%%%%%%%%%%%%%%%%

In \cite{Gundlach20}, we showed the
existence of a two-parameter family of rigidly rotating, stationary star
solutions for any causal equation of state $P=P(\rho)$. These solutions are
analytic everywhere including at the center, and have finite total
mass $M$ and angular momentum $J$. The two free parameters can be taken
to be $\mys$, giving the overall length scale of the star, and $\Omegas$,
parametrizing the rotation of the star. The latter is defined so that
$\Omegas = 0$ corresponds to nonrotating stars.
We can write these exact solutions as
\begin{equation}
\label{2p1_ss_crit:Zstationary}
Z(R) = \check Z\left({R \over \mys}; \mus, \Omegas \right),
\end{equation}
where $\check Z$ is the corresponding exact stationary solutions.
The cosmological constant enters this picture through a specific combination,
parametrized by the dimensionless quantity $\mus$,
\begin{equation}
\mus := -\Lambda \mys^2-\Omegas^2.
\end{equation}
In what follows, all quantities pertaining to the exact stationary
solutions have a check symbol, as in \eqref{2p1_ss_crit:Zstationary}.

The parameters $\mus$ and $\Omegas$ correspond to $\mu$ and $\Omega$
in \cite{Gundlach20}. (The tilde is used to distinguish $\Omegas$ from
the unrelated conserved variable $\Omega$). Regular stationary
solutions exist only for $0 \leq \mus \leq 1$, and $\mus$ can be
interpreted as parametrizing the competition of the attractive
acceleration induced by a negative cosmological constant and the
centrifugal acceleration.

\begin{figure*} \centering
	\includegraphics[width=0.49\textwidth]{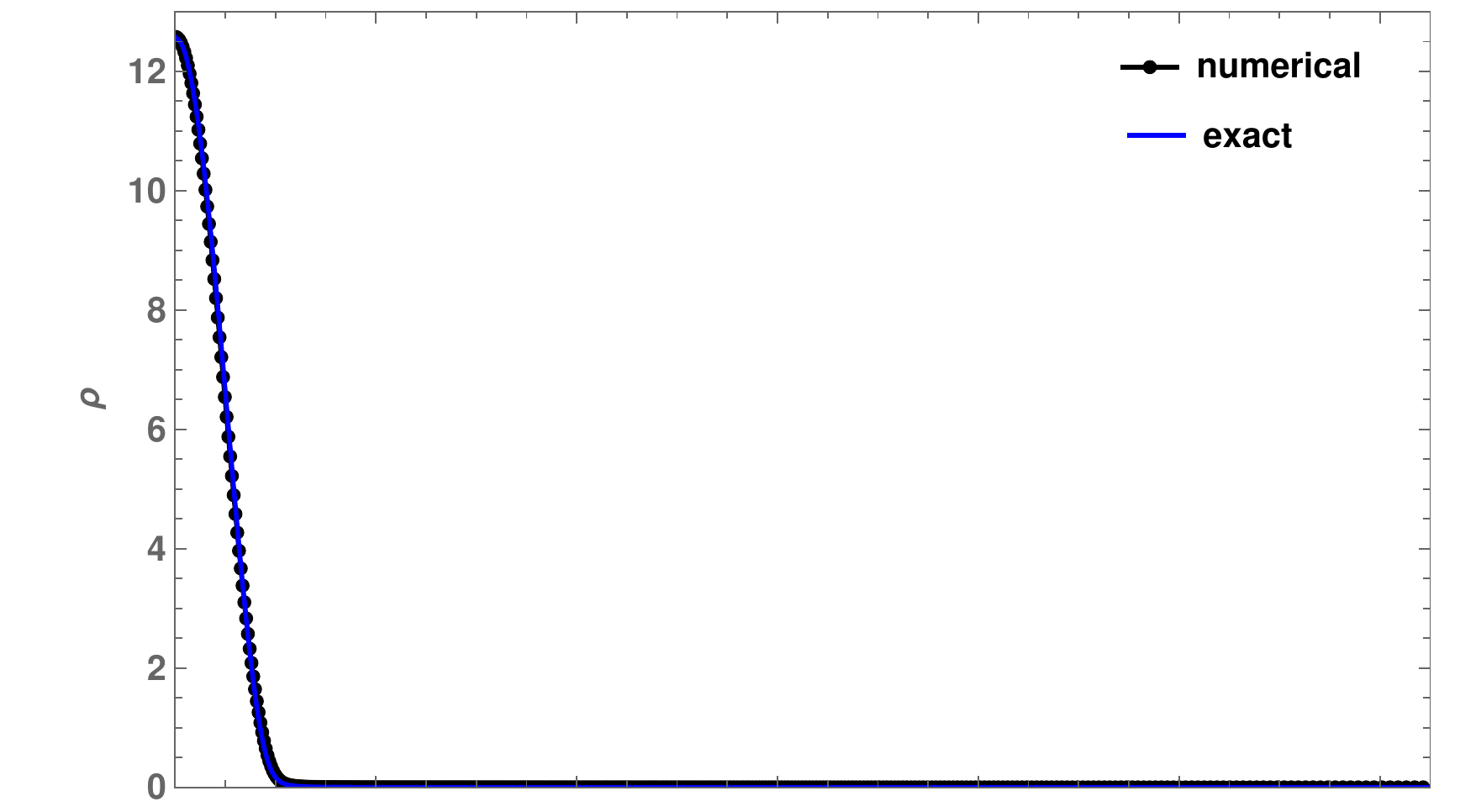}
	\includegraphics[width=0.49\textwidth]{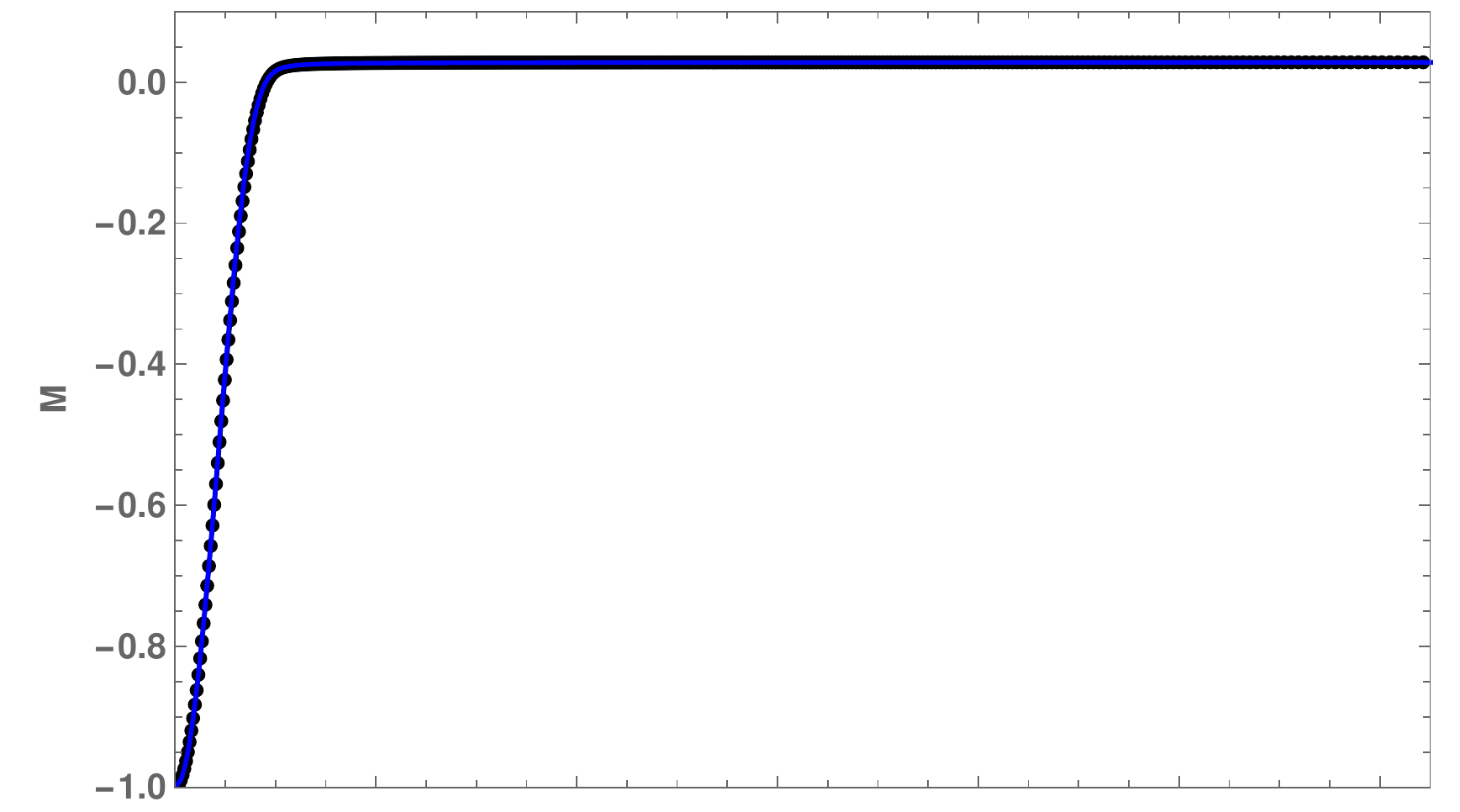}
	\includegraphics[width=0.49\textwidth]{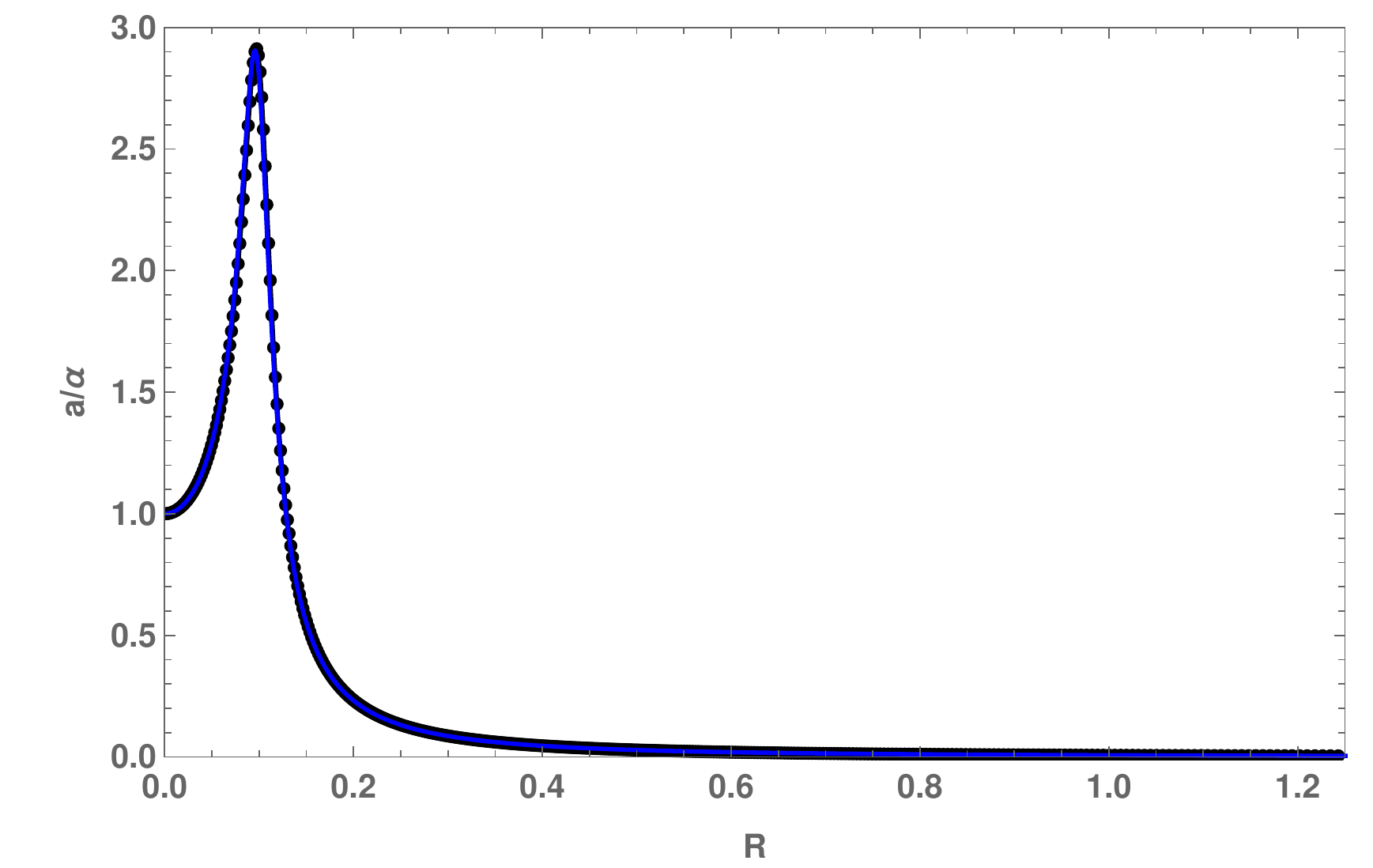}
	\includegraphics[width=0.49\textwidth]{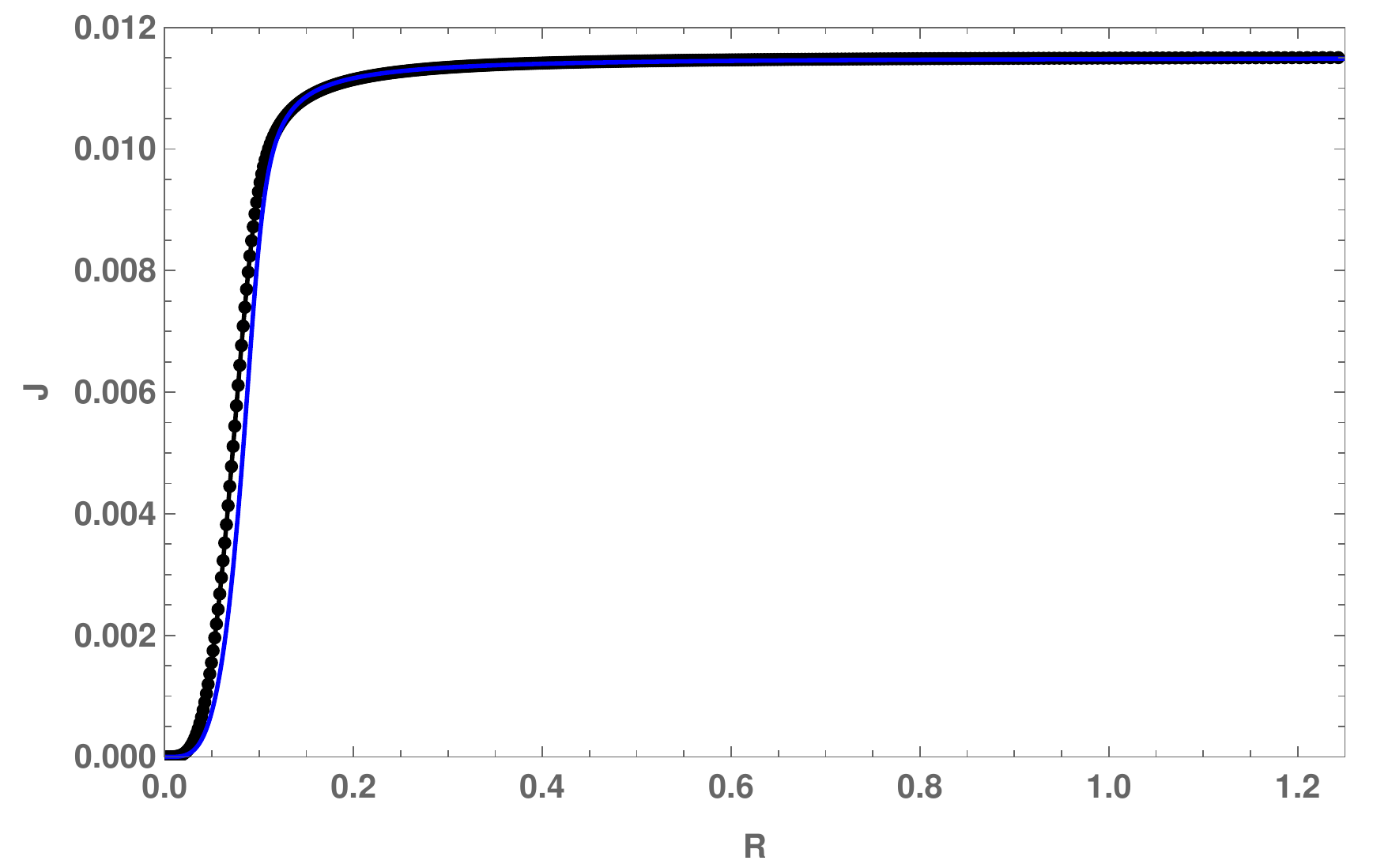}
	\caption{Comparison of the numerical solution for our best
		subcritical data during the critical regime (black dotted
		lines) with the exact stationary solutions (solid blue
		lines). For the numerical data, we chose here $\kappa=0.4$
		with $p_\omega=1.0$.}
	\label{fig:typeI_crit_sol}
\end{figure*}

In Fig.~\ref{fig:typeI_crit_sol}, we compare our best subcritical
numerical solution at $t=2.1$, where the solution is approximately
stationary, to the family of stationary solutions. Since the
stationary solutions form a two-parameter family, we need to fit those
two parameters. For this, we match the central density and total
angular momentum, using the angular momentum at the numerical outer
boundary as a proxy for the total angular momentum. The
matching conditions are
\begin{eqnarray}
J_\text{OB} (1-\kappa) &\equiv& 4 \kappa \ell (1-\mus) \Omegas
\sqrt{\mus+\Omegas^2}, \label{JOB_stationary}\\ 1-\mus &\equiv& 8 \pi
\kappa \rho_0 \ell^2 (\mus+\Omegas^2).
\end{eqnarray}
Note that there is a slight abuse of language here; the total angular
momentum (and mass) of the system is a conserved quantity. The angular
momentum (and mass) at the numerical outer boundary (at finite radius)
differ from it, since a bit of spin (and mass) are radiated away
through this boundary at the start of the evolution, when the solution
has yet to enter the critical regime. At late times, the
approximately time-independent value of the angular momentum (and
mass) at the outer boundary during the critical regime is a
fixed fraction of the constant total spin (and mass)
of the system.

Our matching of the angular momentum therefore introduces a small
systematic error. At first, we attempted to ask for the value of $R_M$
between the numerical and exact solutions to agree. This is also
possible, but $R_M$ depends very weakly on rotation. Such a
matching condition is therefore very sensitive to numerical error.

In the example of Fig.~\ref{fig:typeI_crit_sol}, our matching
procedure gives $\mys \simeq 0.0882$ and $\Omegas \simeq
0.0498$. We then find very good agreement for $\rho$, $M$ and
$a/\alpha$ at all $R$. We find slightly less good agreement for the
angular momentum. Since $J$ is relatively small, it is likely that
this is a numerical error, although its precise nature is unknown.

% Fig 3 %%%%%%%%%%%%%%%%%%%%%%%%%%%%%%%%%%%%%%%%%%%%%%%%%%%%%%%%%%%%%%%%%%
%%%%%%%%%%%%%%%%%%%%%%%%%%%%%%%%%%%%%%%%%%%%%%%%%%%%%%%%%%%%%%%%%%%%%%%%%

We remark here that, at first, we attempted to push the numerical
outer boundary further out. We noticed however, for $R_\text{max}
\gtrsim 4$, there is a clear numerical error that builds up and causes
mass and spin to slowly radiate away. For example, for $R_\text{max}
\simeq 1.25$, the total mass and angular momentum stay constant in the
critical regime within $\ll 1\%$ (see
Fig.~\ref{fig:RM_rho0_MOB_JOB}). On the other hand, for
$R_\text{max}\simeq 4$, they decrease slowly and approximately
linearly. As compared to the near constant value that they take for
$R_\text{max} \simeq 1.25$, they further decrease by about $29\%$ and
$16\%$ for $R_\text{max} \simeq 4$, respectively. This is measured up
until the time when the ``star'' disperses. This decay rate is also
different for larger $R_\text{max}$. We do not have an explanation for
this numerical error.

%%%%%%%%%%%%%%%%%%%%%%%%%%%%%%%%%%%%%%
\subsection{$\kappa\gtrsim 0.43$: Type~II critical collapse}
%%%%%%%%%%%%%%%%%%%%%%%%%%%%%%%%%%%%%%

%%%%%%%%%%%%%%%%%%%%%%%%%%%%%%%%%%%%%%
\subsubsection{Curvature and mass scaling}
%%%%%%%%%%%%%%%%%%%%%%%%%%%%%%%%%%%%%%

In spherical symmetry, we have given strong evidence that the critical
phenomena are type~II for $\kappa \geq 0.43$ (the apparent-horizon
mass and curvature scale as some power law and the critical solution
shrinks quasistatically to arbitrarily small size); see
\cite{Bourg21}. As we did for $\kappa \leq 0.42$, we now consider
families of initial data with different initial angular momentum
$p_\omega$. In the following, we focus our attention on the case
$\kappa=0.5$.

\begin{figure*}
	\includegraphics[width=2.\columnwidth, height=1.6\columnwidth]
		{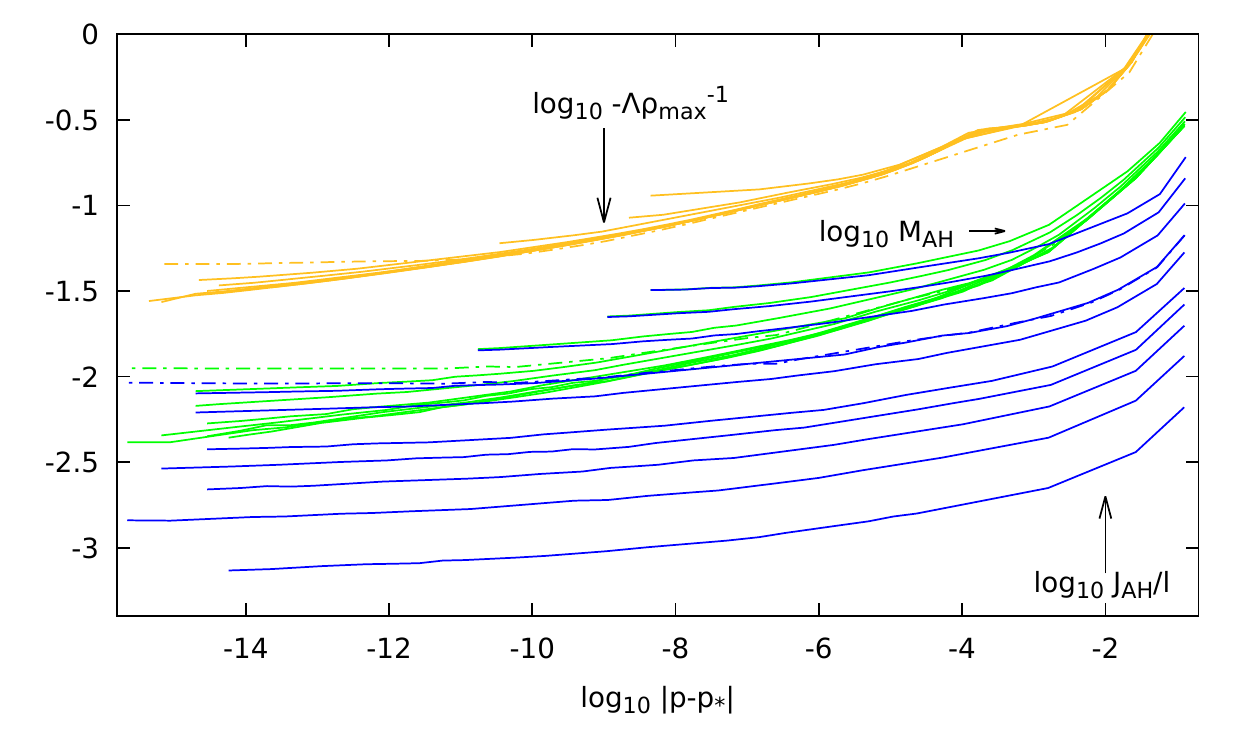}
	\caption{Log-log plot demonstrating the power-law scaling of
		$-\Lambda \rho_\text{max}^{-1}$ (orange, upper group of
		curves), $M_\text{AH}$ (green, middle group of curves) and
		$J_\text{AH}/\ell$ (blue, lower group of curves) for
		different values of $p_w$. Solid lines correspond to
		centered initial data, using the Godunov
		limiter. $J_\text{AH}$ decays more slowly than
		$M_\text{AH}$. For small spin-to-mass ratio, we find typical
		type~II behavior, while for larger spin-to-mass ratio, the
		scaling smoothly flattens as extremality is
		approached. Dashed-dotted lines show an evolution with
		large initial spin ($p_\omega=0.2$), but using the MC
		limiter. There, the spin-to-mass ratio completely flattens
		beyond a certain level of fine-tuning, but this is a
		numerical artifact.}
	\label{fig:typeII_powerlaw}
\end{figure*}

In Fig.~\ref{fig:typeII_powerlaw}, we plot $-\Lambda
\rho_\text{max}^{-1}$ (orange group of curves), $M_{AH}$ (green group
of curves) and $J_{AH}/\ell$ (blue group of curves) as a function of
$p-p_\star$ for centered initial data and with different $p_w$, namely,
$p_w = 0.02$, $0.04$, $0.06$, $0.08$, $0.1$, $0.16$, $0.2$, $0.3$,
$0.4$ and $0.5$ (solid lines). The dashed-dotted lines were obtained
with $p_w = 0.2$, but using the monotonized central (MC) limiter.

This plot illustrates multiple points. First, for ``small'' initial
spin, we find typical type~II phenomena where the density, mass and
spin scale like power laws. Second, the angular momentum $J_\text{AH}$
scales \textit{more slowly} than $M_\text{AH}$. In particular, the
spin-to-mass ratio of the resulting black hole \textit{increases} as
we fine-tune to the black-hole threshold; see also
Fig.~\ref{fig:MJ_manifold_time}. For larger initial spin such as $p_w
\geq 0.2$ (or respectively more fine-tuning for smaller $p_\omega$),
the spin-to-mass ratio approaches extremality. It does not go beyond,
however, because the critical solution becomes stationary and, as a
result, the power-law scaling with respect to $p-p_*$ also smoothly
levels off.

The phenomena highlighted above have also been checked to hold for the
off-centered and initially ingoing data, although we chose not to
include them in the plot to avoid cluttering.

Note that for the evolution using the MC limiter (dashed-dotted), the
scaling stops abruptly and the spin-to-mass ratio seems to remain
constant with further fine-tuning. For reasons that will be made
clearer later, this is expected to be a numerical artifact: for
relatively large spin-to-mass ratio, numerical instabilities are
generated at the numerical outer boundary, similar to those we
faced when simulating type~I phenomena in the spherically symmetric
case. These instabilities travel inward and produce shocks near the
center, which are absent using a more dissipative limiter such as the
first-order Godunov limiter.

% Fig 4 %%%%%%%%%%%%%%%%%%%%%%%%%%%%%%%%%%%%%%%%%%%%%%%%%%%%%%%%%%%%%%%%%
%%%%%%%%%%%%%%%%%%%%%%%%%%%%%%%%%%%%%%%%%%%%%%%%%%%%%%%%%%%%%%%%%%%%%%%%

\begin{figure*}
	\includegraphics[width=1.\columnwidth, height=1.\columnwidth]
		{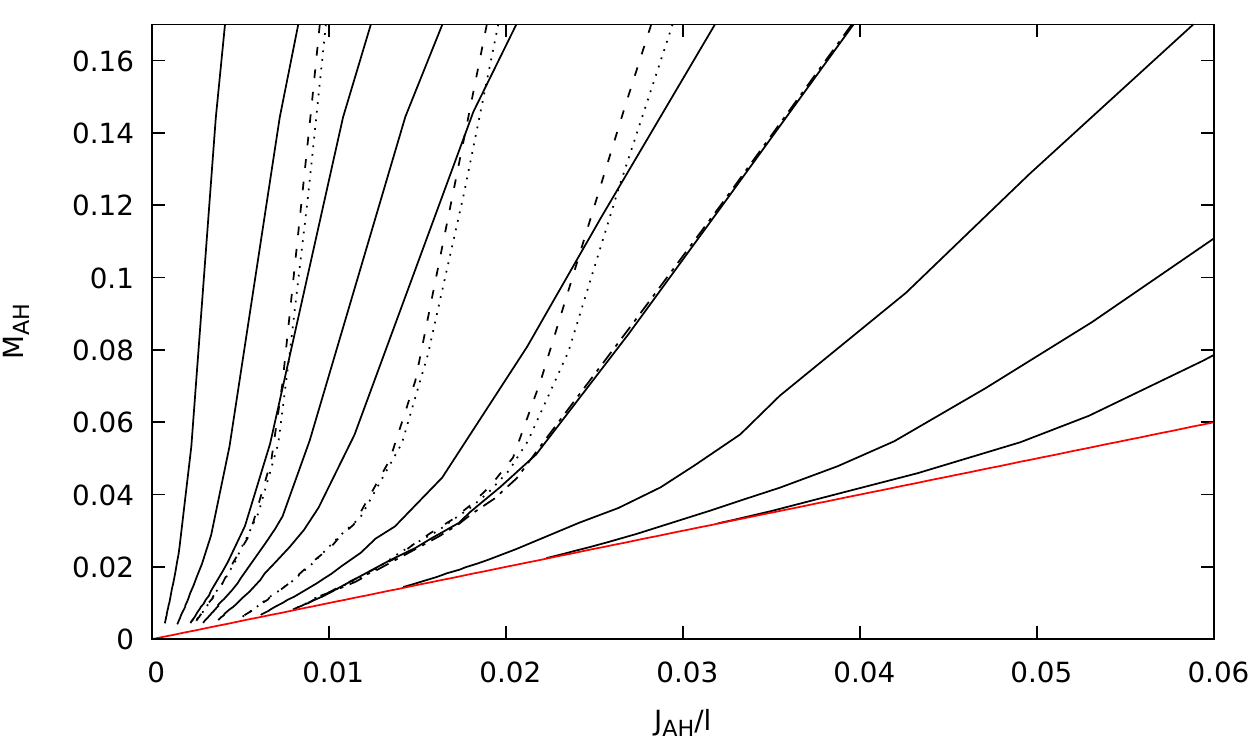}
	\includegraphics[width=1.\columnwidth, height=1.\columnwidth]
		{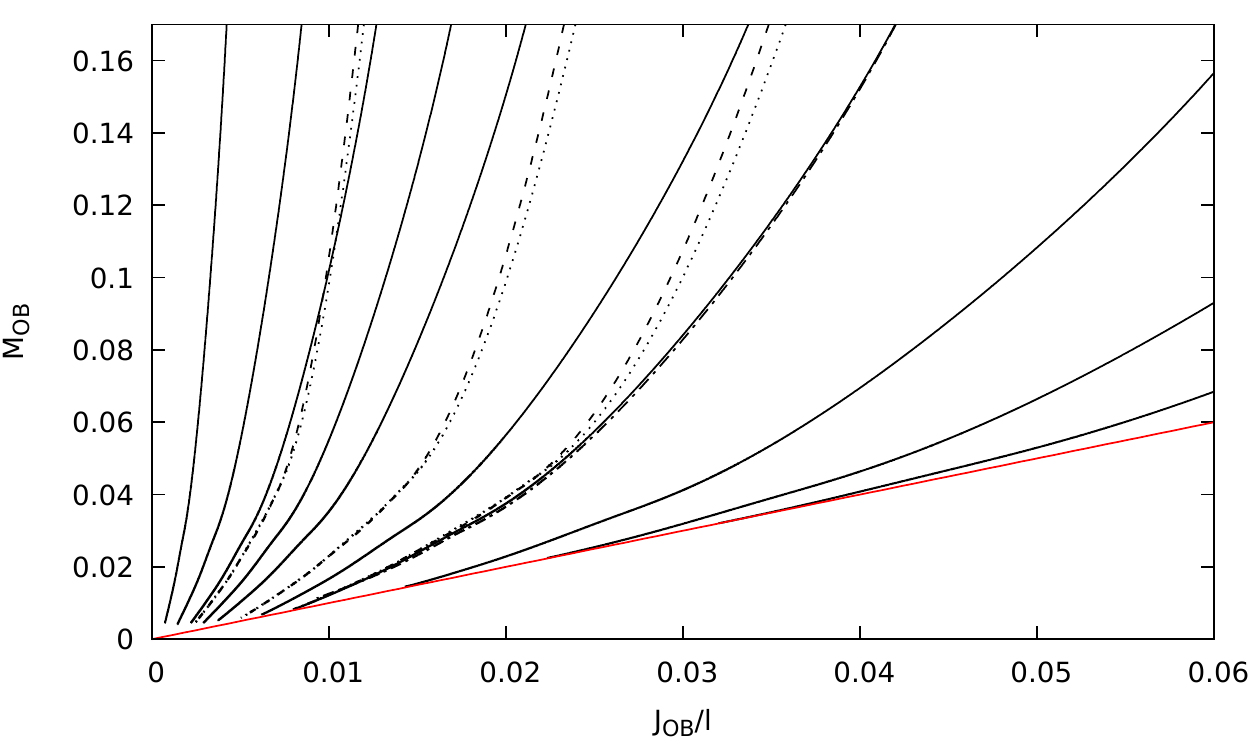}
	\caption{Left plot: parametric plot of the trajectories of
		$M_\text{AH}$ and $J_\text{AH}$ in
		Fig.~\ref{fig:typeII_powerlaw} (solid and dashed-dotted
		lines), in the $M_\text{AH}$-$J_\text{AH}$ plane. The
		parameter along each curve is $p-p_*$. We have also
		added the trajectories of off-centered (dotted lines) and
		ingoing (dashed lines) initial data for three different
		values of $p_w$. The red line corresponds to
		$J_\text{AH} = M_\text{AH}\ell$. Right plot: a
		similar plot, using the same convention, except that we now
		consider the mass and angular momentum evaluated at the
		numerical outer boundary, for our best supercritical data.
		The parameter along each curve is now time.}
	\label{fig:MJ_manifold}
\end{figure*}

\begin{figure*}
	\includegraphics[width=1.\columnwidth, height=1.\columnwidth]
		{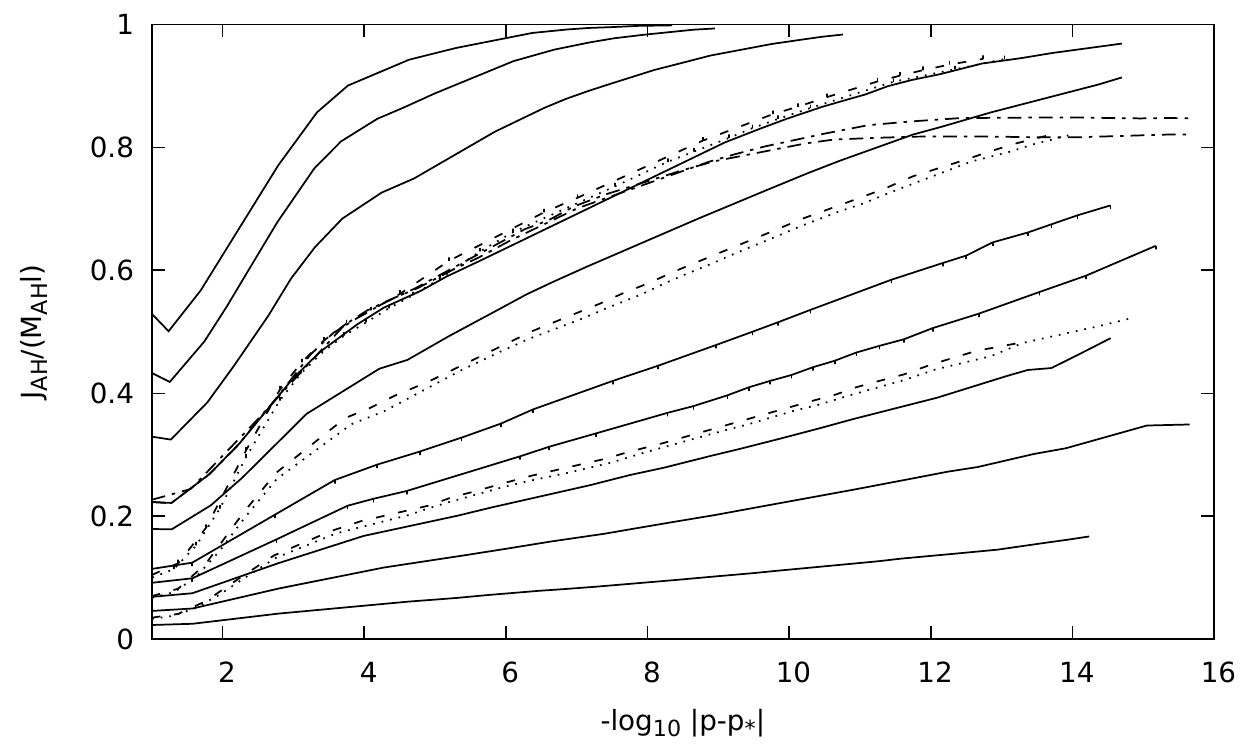}
	\includegraphics[width=1.\columnwidth, height=1.\columnwidth]
		{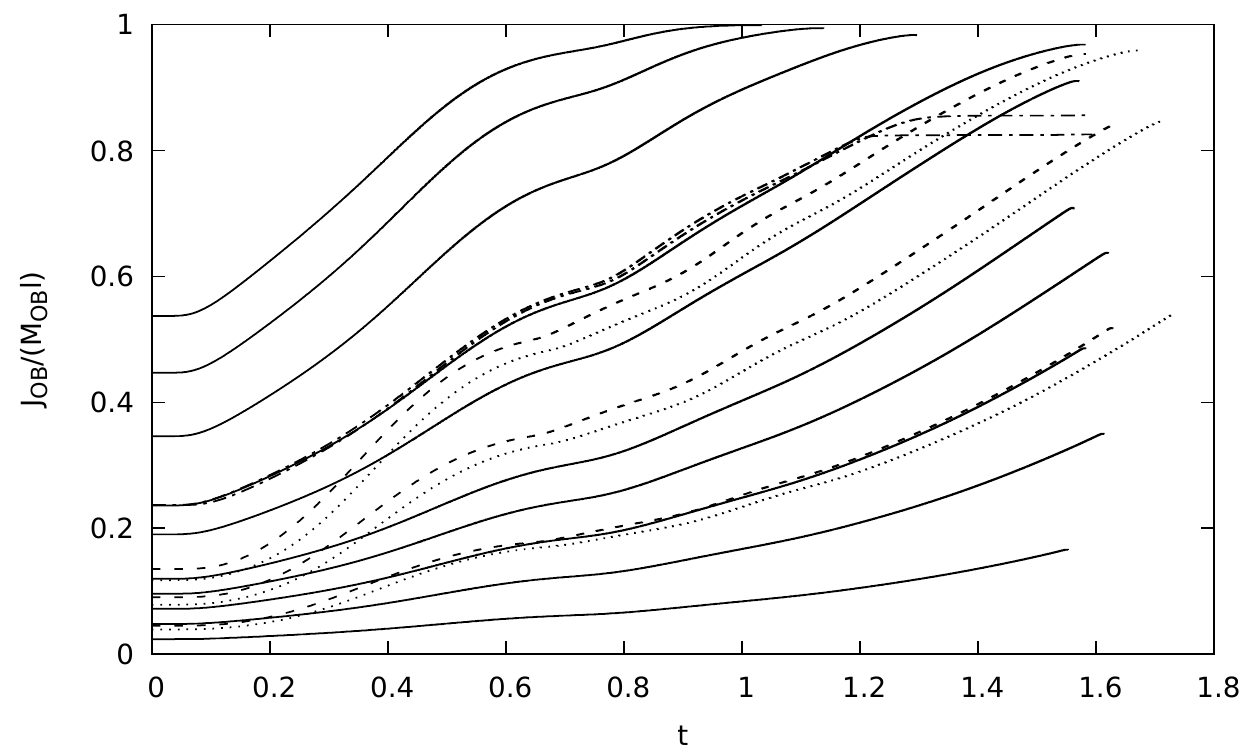}
	\caption{Left plot: log-log plot of $J_\text{AH} \over
		M_\text{AH} \ell$ against $p-p_\star$, for the same data as
		in Fig.~\ref{fig:MJ_manifold}, with the addition of a second
		dashed-dotted curve using the MC limiter but with a larger
		grid. The spin-to-mass ratio approaches extremality as we
		fine-tune to the black-hole threshold, except for the two MC
		limiter cases. Right plot: parametric plot of $J_\text{OB}
		\over M_\text{OB} \ell$, as a function of time $t$, for our
		best supercritical evolutions. The critical regime
		begins at $t \simeq 0.8$ for centered initial data, and at
		$t \simeq 1.0$ for ingoing and off-centered initial data.}
	\label{fig:MJ_manifold_time}
\end{figure*}

In Fig.~\ref{fig:MJ_manifold}, we explore the relationship between
$J_\text{AH}$ and $M_\text{AH}$ more explicitly, for the critical
solution and the final black hole. On the left, we show a parametric
plot of $M_\text{AH}$ against $J_\text{AH}/\ell$ for the same initial
data as in Fig.~\ref{fig:typeII_powerlaw} (and using the same
convention for the lines). We have added the bisections from
the off-centered and ingoing initial data with $p_w = 0.02$, $0.04$ and
$0.06$ (dotted and dashed lines, respectively). The red line
corresponds to $J_\text{AH} = M_\text{AH} \ell$. The trajectories in
the $(M_\text{AH},J_\text{AH})$ plane are clearly family dependent,
although as we increase the fine-tuning and the black holes become
smaller (the bottom left corner) all the curves become approximately
parallel to each other. This provides some evidence that the critical
phenomena are controlled by a unique one-parameter family of critical
solutions (universality).

On the right, we show a parametric plot of the mass $M_\text{OB}$
against the angular momentum $J_\text{OB}/\ell$, for our best
supercritical data. As expected, both plots are qualitatively very
similar, since one expects the mass and spin at the apparent horizon
to be in some fixed ratio to their values at the numerical outer
boundary, at the time when the solution veers off from the critical
solution. It turns out that, for close to critical data,
$M_\text{AH}\simeq M_\text{OB}$ and $J_\text{AH}\simeq J_\text{OB}$ at
this time, and so this fixed fraction is almost one.

In Fig.~\ref{fig:MJ_manifold_time}, we use again the same data as in
Fig.~\ref{fig:MJ_manifold}. On the left plot, we show the
spin-to-mass ratio $J_\text{AH}/(M_\text{AH} \ell)$ as a function of
$p-p_\star$ for different levels of fine-tuning. The only addition is
a second bisection using a MC limiter (a second dashed-dotted curve),
but where the numerical outer boundary is at $R_\text{max} \simeq 2.65 \simeq 4.17 \ell$.
The fact that both dashed-dotted lines do not level off at the same
spin-to-mass ratio gives evidence that the unphysical numerical outer
boundary spoils the numerical results in that highly rotating
regime. Instead, the results from the Godunov limiter, which removes
the aforementioned instabilities, are more plausible; in the limit of
fine-tuning, the black hole is extremal.

Similarly, in the right plot, we show the spin-to-mass ratio
evaluated at the numerical outer boundary, for our best supercritical
data, against proper time $t$. Both left and right plots are
qualitatively similar. This is expected since the more fine-tuned the
initial data are to the black-hole threshold, the longer it takes for
the growing mode to dominate the perturbation, and so the more the
spin-to-mass ratio can grow, before the solution collapses.

% Fig 5 %%%%%%%%%%%%%%%%%%%%%%%%%%%%%%%%%%%%%%%%%%%%%%%%%%%%%%%%%%%%%%%%%
%%%%%%%%%%%%%%%%%%%%%%%%%%%%%%%%%%%%%%%%%%%%%%%%%%%%%%%%%%%%%%%%%%%%%%%%

If we define the time where the critical regime starts to be the time
from which the central density shows critical scaling (at least for small
rotation), then this occurs at $t \simeq 0.8$ for centered initial data
(see, for example, Fig.~\ref{fig:RM_rho0_MOB_JOB}), while it occurs at
$t \simeq 1.0$ for ingoing and off-centered initial data.
In the right plot of Fig.~\ref{fig:MJ_manifold_time}, the
different trajectories do not cross during this critical regime. This
again provides evidence that, in a suitable adiabatic limit, a given
pair $(M_\text{AH}, J_\text{AH})$ in a quasistationary sequence
uniquely determines its evolution.

% Fig 6 %%%%%%%%%%%%%%%%%%%%%%%%%%%%%%%%%%%%%%%%%%%%%%%%%%%%%%%%%%%%%%%%%
%%%%%%%%%%%%%%%%%%%%%%%%%%%%%%%%%%%%%%%%%%%%%%%%%%%%%%%%%%%%%%%%%%%%%%%%

%%%%%%%%%%%%%%%%%%%%%%%%%%%%%%%%%%%%%%
\subsubsection{The critical solution}
%%%%%%%%%%%%%%%%%%%%%%%%%%%%%%%%%%%%%%

We have seen in Fig.~\ref{fig:MJ_manifold} that for sufficient
fine-tuning (on the left) or at sufficiently late times (on the
right), the trajectories in the $J$-$M$ plane do not cross,
providing evidence that there is a universal one-parameter family of
critical solutions that fibrates the $J$-$M$ plane. Independently,
the fact that we can make arbitrarily small black holes or
arbitrarily large central densities by fine-tuning generic
one-parameter families of initial data suggests that each member of
this family of critical solutions has precisely one unstable
perturbation mode.

As in the spherically symmetric case, the profile of the
solution can be roughly split into two regions: one central region
where most of the density lies and whose size shrinks in time and an
outer region (``atmosphere'') where the mass and spin are
approximately constant in space and decrease to zero (exponentially)
in time.

At a stage where the effect of spin can still be regarded as
perturbative, one would expect the critical solution to be
approximated by the quasistatic solution in spherical symmetry, plus
a (unique, growing) nonspherical perturbation proportional to
$p_w$ that carries the angular momentum. In particular, one would
expect that the solution shrinks exponentially quickly in time.

\begin{figure*}
	\includegraphics[width=1.\columnwidth, height=1.\columnwidth]
		{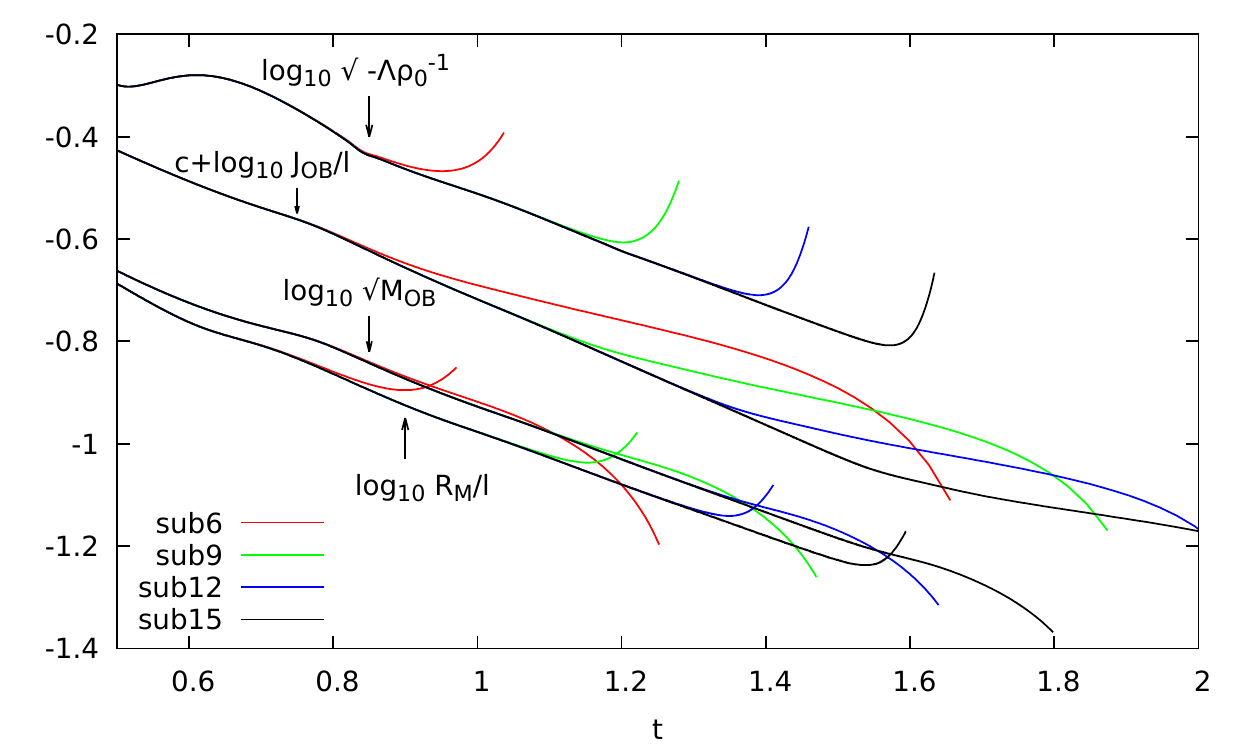}
	\includegraphics[width=1.\columnwidth, height=1.\columnwidth]
		{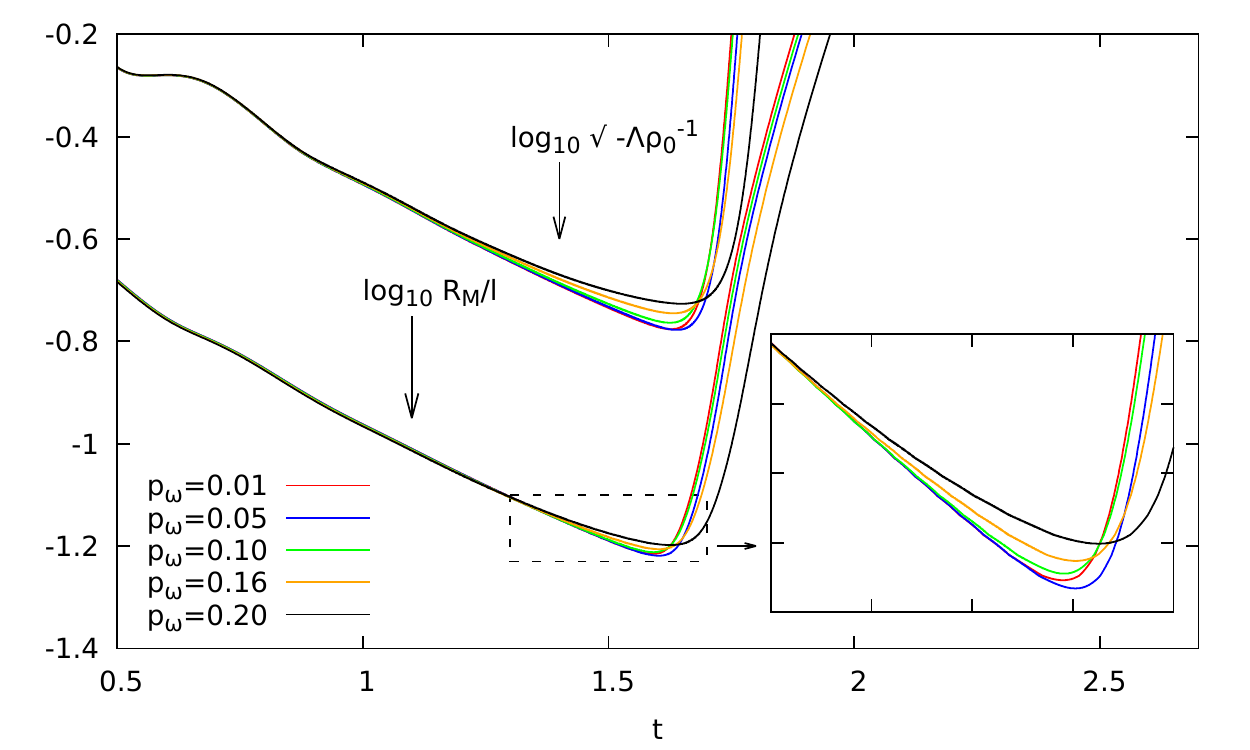}
	\caption{Left plot: lin-log plot of $R_M(t)/\ell$,
		$\sqrt{-\Lambda \rho_0^{-1}(t)}$, $\sqrt{M_\text{OB}(t)}$,
		and $J_\text{OB}(t)/\ell$ for sub6 to sub15 centered initial
		data with $p_w=0.01$, using the MC limiter. We observe that,
		as we fine-tune to the black-hole threshold, the solution
		approaches an intermediate attractor solution in which
		$R_M$, $\rho_0^{-1}$, $M_\text{OB}$ and $J_\text{OB}$
		decrease exponentially. Less fine-tuned initial data peel
		off from this critical line sooner than more fine-tuned
		data, leading to critical scaling of the maximum density,
		etc. Note that $J_\text{OB}$ decreases more slowly than
		$M_\text{OB}$, as is the case for their values at the
		apparent horizon. There is a second regime for
		$J_\text{OB}$, after the critical regime, where it takes on
		another, less pronounced, power-law scaling. Right plot:
		lin-log plot of $\sqrt{-\Lambda \rho_0^{-1}(t)}$ and
		$\sqrt{R_M(t)/\ell}$ for our best subcritical (sub15) data
		with $p_w=0.01$, $0.05$, $0.1$, $0.16$, and $0.20$ using the
		Godunov limiter. At late times, these quantities decrease
		more slowly for larger $p_w$ and correspondingly larger
		spin-to-mass ratio.}
	\label{fig:RM_rho0_MOB_JOB}
\end{figure*}

To confirm this, we plot, on the left of
Fig.~\ref{fig:RM_rho0_MOB_JOB}, the logarithms of $R_M(t)/\ell$,
$\sqrt{-\Lambda \rho_0^{-1}(t)}$, $\sqrt{M_\text{OB}(t)}$ and
$J_\text{OB}/\ell$ for sub6, sub9, sub12, and sub15 centered data with
$p_w = 0.01$. For those plots, we used the MC limiter, as even at
sub15, the spin-to-mass ratio still remains relatively small, and the
aforementioned instabilities are not present.

We find, as anticipated from our study in spherical symmetry, that
these quantities are exponential functions of $t$. $J_\text{OB}$
scales slower then $M_\text{OB}$ as it is the case for the
apparent-horizon mass and spin, so that the spin-to-mass ratio
$J_\text{OB}/(M_\text{OB}\ell)$ increases. Note that right after the
critical regime, the trajectories for each variable at different levels
of fine-tuning align, up to a rescaling and a shift in time. This
kind of behavior is expected in a spacetime where the cosmological
constant is dynamically irrelevant and where the field equations
become approximately scale invariant.

Note that after the critical regime, $J_\text{OB}$ enters what seems
to be a second regime, still decreasing exponentially, but noticeably
more slowly than during the critical regime.

On the right of Fig.~\ref{fig:RM_rho0_MOB_JOB}, we plot the logarithm of
$R_M(t)/\ell$ and $\sqrt{-\Lambda \rho_0^{-1}(t)}$ for our best subcritical
evolution (sub15) with $p_w=0.01$, $0.05$, $0.1$, $0.16$, $0.2$ using
again centered initial data, but with the Godunov limiter. The solution
first shrinks and displays typical type~II phenomena for ``small''
$p_w$. During this phase, $R_M(t)/\ell$ and $\sqrt{-\Lambda
\rho_0^{-1}(t)}$ are completely independent of $p_\omega$, as we
would expect while angular momentum is a perturbation.

As the spin-to-mass ratio has become sufficiently large, both $R_M$
and $\rho_0^{-1}$ decrease more slowly and start to level off. For the
same fine-tuning, this corresponds to larger values of $p_w$; see the
inset on the right of Fig.~\ref{fig:RM_rho0_MOB_JOB}.
We do not have a satisfactory dynamical explanation for this.

Together with Fig.~\ref{fig:typeII_powerlaw}, one expects that in the
limit of perfect fine-tuning, the critical solution becomes
stationary at late times, and correspondingly, on the black-hole
side of the threshold of collapse, the black hole becomes extremal.

% Fig 7 %%%%%%%%%%%%%%%%%%%%%%%%%%%%%%%%%%%%%%%%%%%%%%%%%%%%%%%%%%%%%%%%%
%%%%%%%%%%%%%%%%%%%%%%%%%%%%%%%%%%%%%%%%%%%%%%%%%%%%%%%%%%%%%%%%%%%%%%%%

In spherical symmetry, the critical solution is quasistatic, meaning that it
adiabatically goes through the sequence of static solutions, with $\mys$
now a function of $t$, and $|\dot \mys(t)|\ll 1$.
Furthermore, $\mus \sim \mys^2$, and it follows that in the
quasistatic ansatz, $\mus(t) \sim \mys(t)^2$. Since the solution shrinks
exponentially, we have the following ansatz for $\mys(t)$:
\begin{equation}
\mys(t) \equiv s_0 e^{-\nu {t \over \ell}}, \label{stansatz}
\end{equation}
for constants $s_0$ and $\nu$.
To leading order in $\dot \mys$, the quasistatic ansatz then takes the form
\begin{equation}
M(t,R) \simeq \check M \left({R \over \mys(t)}; \mus(t), \Omegas=0\right),
\end{equation}
and similarly for other suitably rescaled variables.

As we hinted before, one can expect, at least in the regime where the
effects of angular momentum are still perturbative, that the solution
can be thought to shrink adiabatically to zero size, going through a
sequence of \textit{stationary} solutions. The picture in spherical
symmetry can then be straightforwardly generalized to axisymmetric
initial data. Specifically, we assume that, to leading order in
$\dot{\mys}$, the critical solution can be well approximated by
\begin{equation}
M(t,R) \simeq \check M \left({R \over \mys(t)}; \mus(t), \Omegas(t)\right),
\label{M_stationary}
\end{equation}
and so on. From the exponential form of $\mys(t)$, we further make a
relatively agnostic ansatz for $\Omegas$ of the form
\begin{equation}
\Omegas(t) =: \Omegas_0 e^{- \phi {t \over \ell}}, \label{Omt}
\end{equation}
where $\Omegas_0$ and $\phi$ are constants. This exponential ansatz
is justified from the fact that, for the family of stationary
solutions, the angular momentum at infinity satisfies
\eqref{JOB_stationary}. In particular $J \sim \Omegas$. From
Fig.~\ref{fig:RM_rho0_MOB_JOB}, $J_\text{OB}$, seen as a proxy for the
corresponding angular momentum at infinity, decays exponentially, thus
suggesting the exponential form for $\Omegas$. Together with the
ansatz for $\mys(t)$ \eqref{stansatz}, we have now four parameters to
fit: $s_0$ and $\nu$ (as in spherical symmetry) and $\Omegas_0$ and
$\phi$.

For $p_w=0.01$, where we only observe type~II behavior to our level
of fine-tuning, we choose to fix them by requiring the central density
and spin at infinity to match those of the stationary solutions at
times $t \simeq 1.0$ and $1.4$, where we believe that the solution is
in its critical regime; see Fig.~\ref{fig:RM_rho0_MOB_JOB}. For the
spin at infinity, we take the spin at the numerical outer boundary as
a proxy. This is justified because in the atmosphere, where the
density is small, the mass and spin are approximately constant in
space. We find the following values:
\begin{eqnarray}
s_0 &\simeq& 0.1913, \qquad \nu \simeq 0.7925, \nonumber\\
\Omegas_0 &\simeq& 0.002110, \quad \phi \simeq 0.1144. \label{s0nuOmphi}
\end{eqnarray}
As expected, the above values of $s_0$ and $\nu$ are consistent with
their values in the spherically symmetric case $p_w=0$. We
have repeated this procedure for the ingoing and off-centered
initial data as well, and summarize the values of $\nu$ and $\phi$ in
Table~\ref{table:nu_phi}. Due to the smallness of $\phi$, the
relative variation of $\phi$ is important. In particular,
it is difficult to confidently say if those variations are purely a
numerical error. However, because Figs.~\ref{fig:MJ_manifold} and
\ref{fig:MJ_manifold_time} suggest some universality for the critical
solution, we are inclined to believe it is so. That is, we
believe that $\phi$, just as $\nu$, is family independent.
We find good agreement between the numerical and exact solutions.

% Table 1 %%%%%%%%%%%%%%%%%%%%%%%%%%%%%%%%%%%%%%%%%%%%%%%%%%%%%%%%%%%%%%%
\begin{table} \centering
	\begin{tabular}{{p{0.6\columnwidth}>{\centering}
		p{0.2\columnwidth}>{\centering\arraybackslash}p{0.15\columnwidth}}}
		\hline \hline
		Initial data ($\kappa=0.5$) & $\nu$ & $\phi$ \\
		\hline
		Off-centered & 0.7844 & 0.0854 \\
		Centered	 & 0.7925 & 0.1144 \\
		Ingoing		 & 0.8049 & 0.0671 \\
		\hline \hline
	\end{tabular}
	\caption{The values of $\nu$ and $\phi$ for different families
		of initial data.}
	\label{table:nu_phi}
\end{table}
%%%%%%%%%%%%%%%%%%%%%%%%%%%%%%%%%%%%%%%%%%%%%%%%%%%%%%%%%%%%%%%%%%%%%%%%

\begin{figure*}
	\includegraphics[width=2.\columnwidth, height=1.4\columnwidth]
		{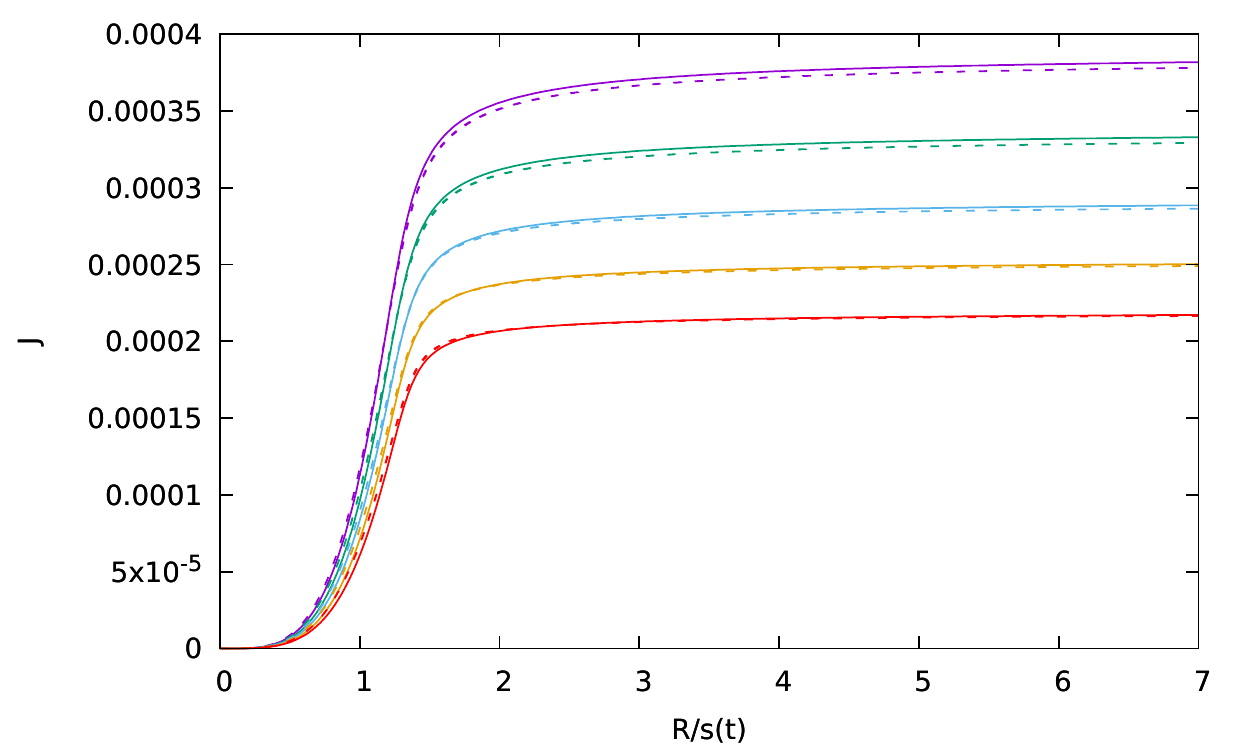}
	\caption{Numerical solution (solid lines) for $J$,
		plotted against $R/s(t)$, at different times during the
		critical regime, $t \simeq 1$, $1.1$, $1.2$, $1.3$ and $1.4$
		for our best subcritical, centered initial data with
		$p_w=0.01$. For comparison, we plot the leading-order term
		of the quasistationary solution for $J = \check J(R/s(t);
		\mus(t), \Omegas(t))$ (dashed lines). We consider
		the ansatz $s(t) = s_0 e^{-\nu t/\ell}$ and $\Omegas(t) =
		\Omegas_0 e^{-\phi t/\ell}$, where the parameters $s_0$,
		$\nu$, $\Omegas_0$, and $\phi$ are given in
		Eq.~\eqref{s0nuOmphi}.}
	\label{fig:crit_sol_J_comparison}
\end{figure*}

In Fig.~\ref{fig:crit_sol_J_comparison}, we compare the leading-order
quasistationary solution for $J$, of the same form as in
Eq.~\eqref{M_stationary} (dotted lines), with the numerical data
(solid lines) at constant time intervals $t \simeq 1$, $1.1$, $1.2$,
$1.3$ and $1.4$.

% Fig 8 %%%%%%%%%%%%%%%%%%%%%%%%%%%%%%%%%%%%%%%%%%%%%%%%%%%%%%%%%%%%%%%%%
%%%%%%%%%%%%%%%%%%%%%%%%%%%%%%%%%%%%%%%%%%%%%%%%%%%%%%%%%%%%%%%%%%%%%%%%

Returning back to the left plot of Fig.~\ref{fig:RM_rho0_MOB_JOB}, the
quasistationarity of the critical solution can be used to explain the
slope of $J_\text{OB}$. Specifically, from \eqref{JOB_stationary}, we
have
\begin{equation}
J_\text{OB} \sim \Omegas s.
\end{equation}	
In the spherically symmetric case, we showed that $M_\text{OB} \sim s^2$. At
the level where angular momentum is still viewed as a perturbation, this is
still expected to hold, so that
\begin{equation}
J_\text{OB}^2 \sim \Omegas^2 M_\text{OB},
\end{equation}
or
\begin{equation}
J_\text{OB}^2 \sim M_\text{OB}^{1+\Delta},
\end{equation}
where
\begin{equation}
\Delta = \phi/\nu
\end{equation}
is small. For the centered initial data for example,
\eqref{s0nuOmphi} gives $\Delta \simeq 0.144$.

%%%%%%%%%%%%%%%%%%%%%%%%%%%%%%%%%%%%%%%%%%%%%%%%%%%%%%%%%%%%%
\section{Conclusion}
%%%%%%%%%%%%%%%%%%%%%%%%%%%%%%%%%%%%%%%%%%%%%%%%%%%%%%%%%%%%%

We have generalized our previous study \cite{Bourg21} of perfect fluid
critical collapse in $2+1$ spacetime dimension to rotating initial
data. We have given evidence that, for $\kappa \lesssim 0.42$, the
critical phenomena are type~I, as in spherical symmetry. The critical
solution is stationary and agrees well with the family of exact
stationary solutions studied in another paper \cite{Gundlach20}. 

The situation for $\kappa \gtrsim 0.43$ is more
complicated. We give evidence for the existence of a universal
one-parameter family of critical solutions, which fibrates the region
$|J/\ell|<M$, $M>0$ of the $J$-$M$ plane. In the limit $|J|/(\ell
M)\ll1$, as well as $s/\ell \ll 1$, angular momentum can be
approximated as a linear perturbation of the nonrotating critical
solution. We then expect both $s$ and $J$ to be exponential functions
of $t$, and so $J$ to be a power of $s$.

However, for supercritical data, the angular momentum of the solution
decreases more slowly than its mass as the black-hole threshold is
approached. The spin-to-mass ratio therefore \textit{increases} as we
fine-tune to the black-hole threshold. We gave strong evidence that in
the limit of perfect fine-tuning, the resulting black hole is
extremal.

For small angular momentum, rotation can be treated as a linear
perturbation of spherical symmetry. The universality of the
one-parameter family of critical solutions, with increasing
$|J|/(\ell M)$, then implies the existence of a single unstable
rotating mode of the spherical critical solution. One may then
wonder how superextremality is avoided. We have seen that the
answer is that the critical solution stops shrinking as $|J|/(\ell
M)\to 1$, and so both $J$ and $M$ stop scaling.

Contrast this with the known situation in rotating fluid collapse in
$3+1$ \cite{Baumgarte16,Gundlach18}. There, the existence of a single
growing angular momentum mode is known for $\kappa<1/9$
\cite{Gundlach02}, but it turns out that nonlinear effects make
$J/M^2$ decrease in the critical solution from the start, even for
small $J$ for $\kappa<1/9$.

As $s/\ell\ll 1$, or $M \ll 1$, the physics become approximately
scale invariant, and the cosmological constant becomes essential only
in a boundary layer at the surface of the contracting star. We expect
that in this limit the one-parameter family of critical solutions
degenerates to a single (somewhat singular) critical solution, up to
an overall rescaling.

We have shown that type~I critical collapse is controlled by
rigidly rotating stationary solutions and type~II by an adiabatic
shrinking sequence of such solutions. The matching between our
numerical results and the exact stationary solutions of
\cite{Gundlach20} is very good, even neglecting the adiabatic
contraction. On the flip side, we have not been able to derive
effective adiabatic equations of motion in the space of stationary
solutions (even in the nonrotating case of \cite{Bourg21}), and so
we are unable to derive either the universal time dependence of the
critical solutions or their unique trajectories in the $J$-$M$
plane.

It is fortunate that our critical solutions appear to be exactly
rigidly rotating, as precisely all rigidly rotating stationary
solutions are known in closed form (for arbitrary equation of state)
\cite{Gundlach20,Cataldo04}. It is plausible that the rigid rotation of
the critical solutions is universal, but we have not tested this, as
we have considered only one family of initial rotation profiles (in
which the angular velocity is approximately constant).

Critical phenomena in $2+1$ dimensions primarily serves as toy model which
makes the transition from spherical to axisymmetric initial data, unlike the
$3+1$ dimensional setting, much more tractable. Between the results from the
scalar field case \cite{Jalmuzna15,Jalmuzna17} and the results for the perfect
fluid here, one can see that the effects of angular momentum are generally far
from trivial and share little resemblance with their higher-dimensional
counterparts.

% Acknowledgements %%%%%%%%%%%%%%%%%%%%%%%%%%%%%%%%%%%%%%%%%%
\begin{acknowledgements}
	The authors acknowledge the use of the IRIDIS 4 High Performance
	Computing Facility at the University of Southampton regarding the
	simulations that were performed as part of this work.
	
	Patrick Bourg was supported by an EPSRC Doctoral Training Grant to
	the University of Southampton.
\end{acknowledgements} 
%%%%%%%%%%%%%%%%%%%%%%%%%%%%%%%%%%%%%%%%%%%%%%%%%%%%%%%%%%%%%

%%%%%%%%%%%%%%%%%%%%%%%%%%%%%%%%%%%%%%%%%%%%%%%%%%%%%%%%%%%%%

%%%%%%%%%%%%%%%%%%%%%%%%%%%%%%%%%%%%%%%%%%%%%%%%%%%%%%%%%%%%%%%%%%%%%%
\end{document}